\documentclass{elsart}

 \usepackage{graphicx}

\usepackage{amssymb}

\begin{document}

\begin{frontmatter}



\title{Azimuthal Correlations between Non-Photonic Electrons and Charged Hadrons in p+p Collisions from STAR}


\author{Xiaoyan Lin ({\it for the STAR Collaboration})}


\address{Central China Normal University, Wuhan 430079, China and
University of California, Los Angeles, CA 90095, USA}

\begin{abstract}
We present the preliminary measurement of azimuthal correlations
between non-photonic electrons and charged hadrons in p+p collisions
at $\sqrt{s_{NN}} = 200$ GeV from STAR. The results are compared to
PYTHIA simulations to estimate the relative contributions of $D$ and
$B$ meson semi-leptonic decays to the non-photonic electrons.
\end{abstract}

\end{frontmatter}

Heavy quarks are valuable probes of the dense medium created in
heavy-ion collisions. Their transport dynamics such as flow and
energy loss reflect QCD properties of the dense matter. Recently,
the STAR and PHENIX collaborations have observed a suppression of
non-photonic electrons much larger than predicted in the $p_{T} \sim
4-8$ GeV/c region~\cite{star1}\cite{phenix1}. This observation
challenges theoretical predictions based on energy loss via induced
gluon radiation for heavy quarks~\cite{armesto}\cite{djordjevic}.
The fact that the relative contributions from semi-leptonic decays
of $D$ and $B$ mesons to the measured non-photonic electron spectrum
is currently unknown at RHIC severely restricts the understanding of
the heavy quark energy loss. A nonzero non-photonic electron $v_{2}$
has been measured for $p_{T} < 2.0$ GeV/c, while at higher $p_{T}$
the $v_{2}$ is observed to decrease with $p_T$~\cite{akiba}. The
quantitative understanding of this feature in heavy quark
measurements requires the knowledge of the relative charm and bottom
contributions to non-photonic electrons.

In this paper we will present preliminary results on the measurement
of azimuthal correlations between high-$p_{T}$ non-photonic
electrons and charged hadrons in p+p collisions at 200 GeV from
STAR. This measurement will help distinguish relative $B$ and $D$
decay contributions to non-photonic electrons because of their
different decay kinematics~\cite{lin}. We will compare the results
to PYTHIA simulations in order to estimate $B$ and $D$ decay
contributions to non-photonic electrons as a function of $p_{T}$ for
$p_{T} > 2.0$ GeV/c.

For this analysis we used about 23.4 million p+p events at
$\sqrt{s_{NN}}=200$ GeV with a cut of $-40$ cm $<$ primary vertex
$z$ $<$ $+30$ cm in order to maximize the statistics while keeping
the amount of material in a reasonable level to minimize photon
conversions. To obtain sufficient statistics at high-$p_{T}$, we
developed high tower triggers corresponding to an energy deposition
of at least 2.6 GeV (HT1) and 3.5 GeV (HT2) in a single tower of the
STAR Barrel Electromagnetic Calorimeter (BEMC)~\cite{beddo}. Around
3.32 million HT1 events and 2.27 million HT2 events were used in
this analysis.

Electron identification uses the information from two of the STAR
subsystems, the TPC~\cite{anderson} and the BEMC. The measurement of
the ionization energy loss, $dE/dx$, for charged tracks in the TPC
gas is used to identify electrons in the first stage. Requiring the
$dE/dx$ values of the selected tracks to be near the expected
electron band in the region $p_{T} >$ 2.0 GeV/c rejects a
significant fraction of the hadron background. After extrapolating
the TPC tracks to the BEMC, we require the ratio $p/E$ to be less
than 1.5 using the momentum information from the TPC, $p$, and the
tower energy information from the BEMC, $E$. Electrons will deposit
almost all of their energy in the BEMC while this is not true for
hadrons. Further hadron rejection is provided by the shower maximum
detector (SMD)~\cite{beddo}, which allows us to cut on the shower
size with high spatial resolution. We require the profile of the
electro-magnetic shower to be within the expectation for electrons.
Combining the power of TPC and BEMC, we can achieve an inclusive
electron sample with purity $> 99\%$ in the $p_{T}$ region up to 5.5
GeV/c. The details of the electron identification technique can be
found in reference~\cite{star1}~\cite{dong}.

The physics signal in this analysis is the angular correlation of
charged hadrons and non-photonic electrons. The background is the
angular correlation of charged hadrons and photonic electrons. There
are primarily two types of photonic electron background. One is from
photon conversion in the detector material and the other is from
scalar meson Dalitz decay~\cite{eidelman}. The contributions from
photon conversions and $\pi^{0}$, $\eta$ Dalitz decays can be
reconstructed from an invariant mass calculation of electron pairs.
The electron candidates are combined with tracks passing a very
loose cut on $dE/dx$ around the electron band. A cut of mass $<$ 0.1
GeV/$c^{2}$ rejects around $70\%$ of the photon conversion and
$\pi^{0}$, $\eta$ Dalitz decay electrons. The fraction of background
electrons from other sources is negligible~\cite{star1}.

\begin{figure}[htb]
                 \includegraphics[width=2.9in]{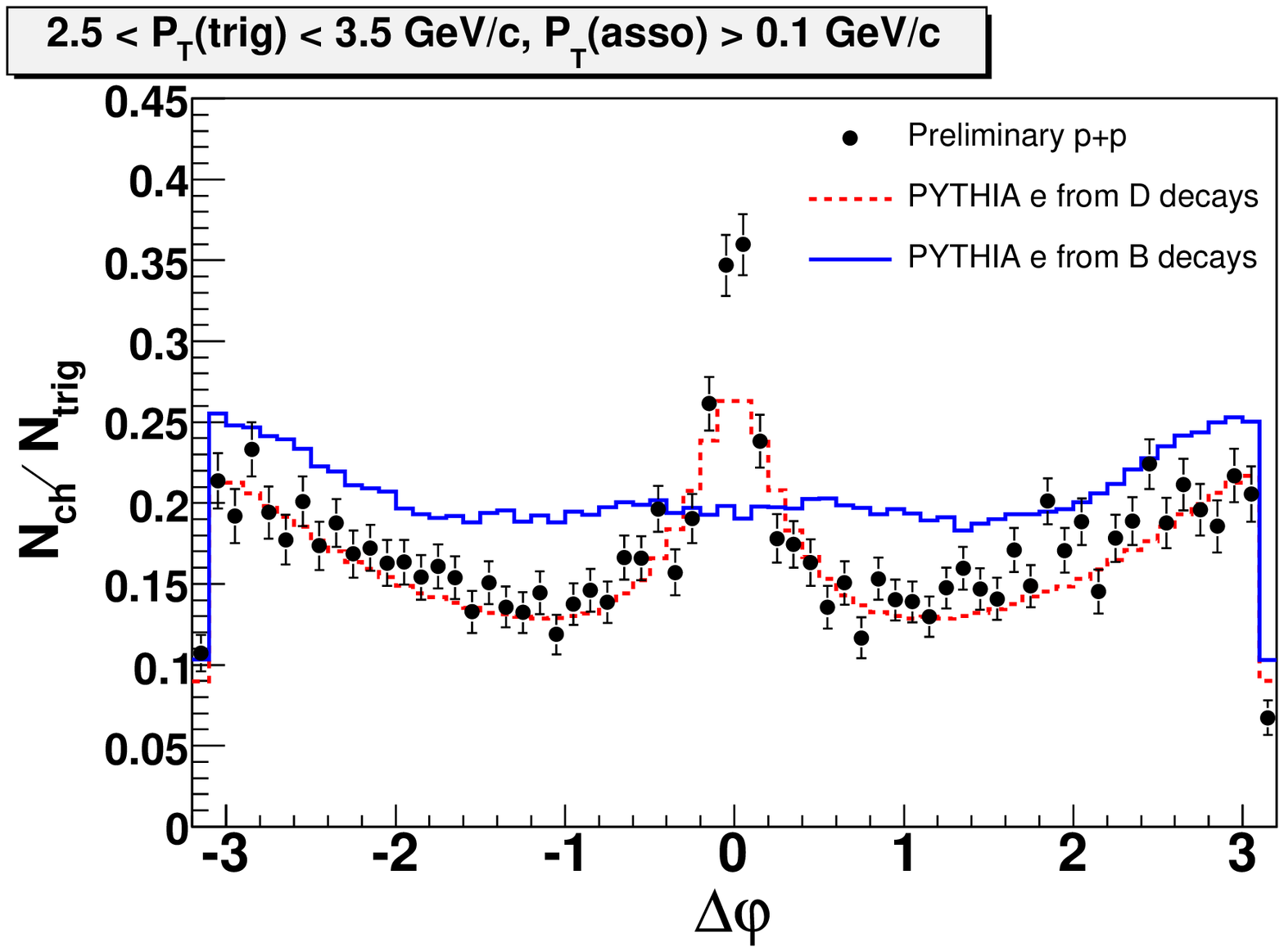}%
                 \includegraphics[width=2.9in]{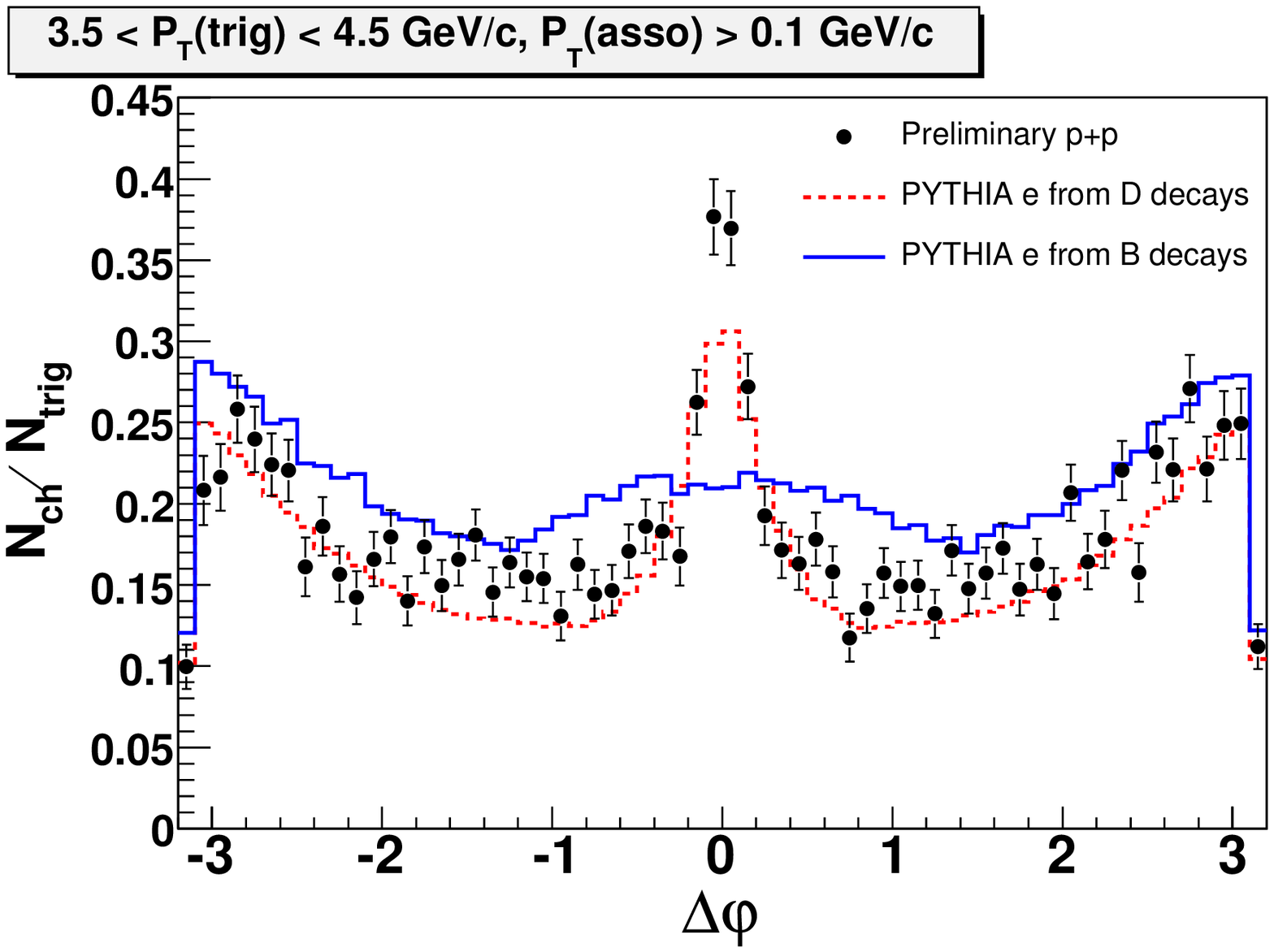}
                 \includegraphics[width=2.9in]{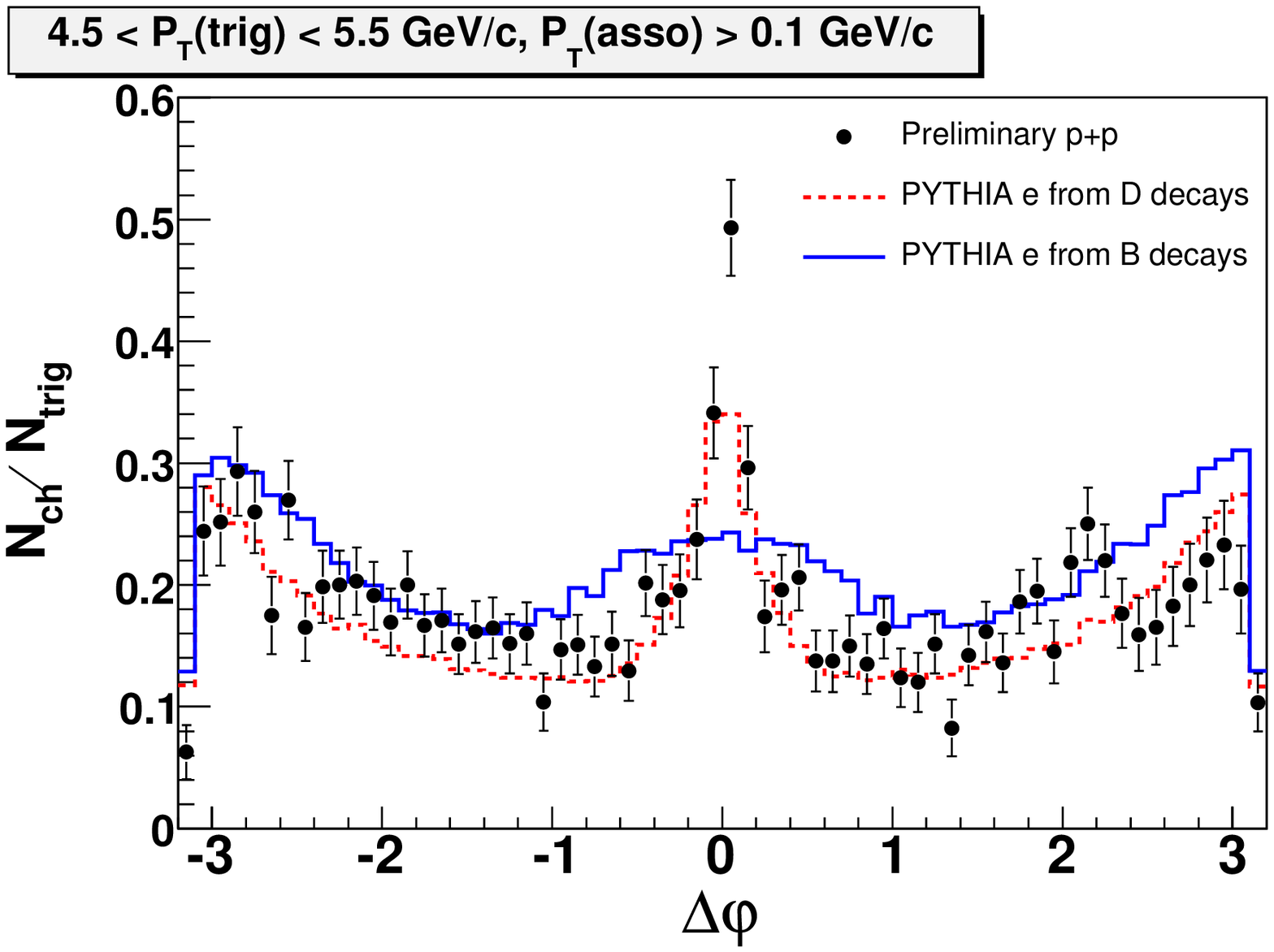}
\caption[]{$\Delta\varphi_{non-pho}$ distributions and the
comparison to PYTHIA simulations for three electron trigger cuts
with associated hadron $p_{T}(assoc)
>$ 0.1 GeV/c. The data is shown as dots, the simulation is depicted by lines. Dashed lines show electrons from $D$ meson decays,
solid lines show electrons from $B$ meson decays.} \label{fig1}
\end{figure}

In order to extract the angular correlation of charged hadrons and
non-photonic electrons, we start with the semi-inclusive electron
sample. The inclusive electron sample includes all tracks that pass
our electron identification cuts. From this sample, we remove those
electrons which satisfy the invariant mass cut where the charges of
the electron pair are opposite ({\it Opp-Sign}). The remaining
electrons form the "semi-inclusive" electron sample. The {\it
Opp-Sign} contains reconstructed-photonic electrons and also
combinatorial background, since non-photonic electrons can be
falsely identified as photonic electrons. This background can be
estimated by the {\it Same-Sign} calculation, which means electrons
satisfy the invariant mass cut where the charges of the electron
pair are the same. The relationship of these samples is: {\it
semi-inc} $=$ {\it inc} $-$ ({\it reco-pho} $+$ {\it combinatorics})
$=$ {\it inc} $-$ ({\it pho} $-$ {\it not-reco-pho} $+$ {\it
combinatorics}) $=$ {\it non-pho} $+$ {\it not-reco-pho} $-$ {\it
combinatorics}. Therefore the signal can be obtained by the
equation: $\Delta\varphi_{non-pho} = \Delta\varphi_{semi-inc} -
\Delta\varphi_{not-reco-pho} + \Delta\varphi_{combinatorics}$.
$\Delta\varphi_{not-reco-pho}$ can be calculated using
$\Delta\varphi_{reco-pho}$ by an efficiency correction after
removing the photonic partner of the reconstructed-photonic
electron. The reason to remove the photonic partner is that for the
reconstructed-photonic electron the photonic partner is found while
for not-reconstructed-photonic electron the partner is missing. The
resulting e-h correlations for reconstructed photonic electrons and
not reconstructed photonic electrons cannot be related to each other
by an efficiency correction factor. Therefore
$\Delta\varphi_{not-reco-pho}$ can be obtained by the equation:
$\Delta\varphi_{not-reco-pho} = (1/\varepsilon -
1)\Delta\varphi_{reco-pho-no-partner}$, where $\varepsilon$ is the
photonic electron reconstruction efficiency and {\it
reco-pho-no-partner} means reconstructed photonic electrons after
removing the photonic partner. The corresponding efficiency can be
calculated from simulations. For simplicity we assume $70\%$ here,
based on previous analyses~\cite{star1}. We still can see a signal
even if the efficiency decreases to $\sim 60\%$. A more detailed
re-evaluation is on progress.

Figure~\ref{fig1} shows $\Delta\varphi_{non-pho}$ distributions and
the comparison to PYTHIA simulations for 2.5 $< p_{T}(trig) <$ 3.5
GeV/c, 3.5 $< p_{T}(trig) <$ 4.5 GeV/c, and 4.5 $< p_{T}(trig) <$
5.5 GeV/c with associated hadron $p_{T} > 0.1$ GeV/c. The data is
shown as dots, the curves are from PYTHIA simulations. The dashed
lines are for electrons from $D$ decays and the solid lines are for
electrons from $B$ decays. Preliminary results can match the
electron-hadron $\Delta\varphi$ distributions from PYTHIA
simulations where electrons come from heavy quark decays only. Our
data indicate that $D$ meson decays are the dominant contribution to
the non-photonic electrons at $p_{T}\sim 2.5-3.5$ GeV/c, and even at
higher $p_{T}$, $p_{T}\sim 4.5-5.5$ GeV/c, $D$ meson decay
contribution is still favored.

In conclusion, the azimuthal correlations between non-photonic
electrons and charged hadrons are a promising tool to distinguish
between $D$ and $B$ decay contributions to non-photonic electrons.
The electron-hadron correlations have been measured in p+p
collisions at $\sqrt{s_{NN}}=200$ GeV from STAR. Preliminary results
show that the $B$ meson decay contribution does not dominate the
measured non-photonic electron yield up to electron $p_{T}$ of 5.5
GeV/c. More quantitative analysis in the future may provide insight
to the puzzle of heavy quark energy loss.




\begin{thebibliography}{00}




\bibitem{star1} J. Adams {\it et al.}, STAR Collaboration, nucl-ex/0607012.
\bibitem{phenix1} S.S. Adler {\it et al.}, PHENIX Collaboration, Phys. Rev. Lett. {\bf 96}, 032301 (2006).
\bibitem{armesto} N. Armesto {\it et al.}, Phys. Rev. D {\bf 71}, 054027 (2005) 054027.
\bibitem{djordjevic} M. Djordjevic {\it et al.}, Phys. Lett. B {\bf 632}, 81 (2006).
\bibitem{akiba} Y. Akiba (for the PHENIX Collaboration),
nucl-ex/0510008.
\bibitem{lin} X.Y. Lin, hep-ph/0602067.
\bibitem{beddo} M. Beddo {\it et al.}, {\it Nucl. Instr. Meth. A} {\bf 499} (2003) 725.
\bibitem{anderson} M. Anderson {\it et al.}, {\it Nucl. Instr. Meth. A} {\bf 499} (2003) 659.
\bibitem{dong} W.J. Dong, Ph.D. thesis, UCLA (2006).
\bibitem{eidelman} S. Eidelman {\it et al.}, {\it Phys. Lett. B} {\bf 592} (2004) $1+$.
\end{thebibliography}
\end{document}